# Analysis of Directional Antenna for Railroad Crossing Safety Applications

Xiaofu Ma[†1], Sayantan Guha[†2], Junsung Choi[†3], Christopher R. Anderson[‡],

Randall Nealy[†4], Jared Withers[⋆], Jeffrey H. Reed[†5], Carl Dietrich[†6]

[†]*Bradley Department of Electrical & Computer Engineering, Virginia Tech, Blacksburg, Virginia, USA*
[xfma[1], sayantg[2], choijs89[3], rnealy[4], reedjh[5], cdietric[6]]@vt.edu

[‡]*Wireless Measurements Group, Department of Electrical & Computer Engineering, U.S. Naval Academy, Annapolis, Maryland, USA*
canderso@usna.edu

[⋆]*Department of Transportation, Federal Railroad Administration, Washington, DC, USA*
jared.withers@dot.gov

*Abstract*—A rapidly deployable and cost-effective railroad crossing early warning system integrated with the railway system is attractive due to its protection of the unmanned grade crossings, which requires a warning system with long-distance communication link. In this paper, we investigate the problem of suitable antenna selection for such a railway warning system. First, the antenna criteria for railroad crossing safety applications are described based on practical system considerations, the safe distances on the road and on the railway. Then, the optimal antenna pattern is derived theoretically to get the smallest size which fits for the practical installation. We also conducted a feasibility study of an array antenna through measurements on a near field scanner[1].

## I. INTRODUCTION

### A. Overview of grade crossing protection

The most common railroad accidents today involve collisions between trains and passenger vehicles at railroad grade crossings [1, 2]. These collisions, due to the size and speed of a train, generally result in significant damage and serious injury, and often even leads to several fatalities. Despite recent efforts by projects such as Operation Lifesaver to install safety features at grade crossings, up to 80% of United States railroad grade crossings are still classified as "unprotected" with no lights, warnings, or crossing gates [1]. Further, for the time period of January – September 2012, nearly 10% of all reported accidents were a result of train-to-vehicle collisions, resulting in nearly 95% of all reported fatalities [2].

Current conventional grade crossing protection systems can be extremely expensive. A typical cost of $ 25,000-$50,000 is required to protect a two-lane grade crossing, while crossings for larger roads can be much more expensive.A system that could provide protection at even $12,500 per crossing would allow railroads to add protection to identified high-risk unprotected crossing at two to four times the current density. However, development of such a low-cost system requires a thorough approach that includes gaining a firm understanding of the operational environment including radio channel characteristics as well as application of safety principles and best practices which focus on railroad industry-specific requirement.

Wireless technologies,such as Dedicated Short Range Communication (DSRC)[3],are currently being explored by the automotive safety industry for Vehicle-to-Vehicle (V2V) and Vehicle-to-Infrastructure (V2I) communications to provide Intelligent Transportation Services (ITS). With features of DSRC, ITSis able to sense and then provide an early warning of a potential collision [4, 5]. This technology can also be adapted for use as a train-to-vehicle collision warning system for unprotected grade crossing[6]. Systems installed onboard on the locomotive would transmit warning messages to approaching vehicles and roadside warning units. Additionally, such systems could provide feedback to train's driver by alerting them aboutpotential incoming traffic.

### B. System deployment framework

The overall schematic of the warning system is shown in Figure 1.The communication system consists three major different nodes: an On-Board Unit (OBU) is lodged on the exterior of the train, a Road-Side Unit (RSU) is at the crossing and OBUsare also installed on the vehicles to which the warning messages are directed. Each of these are DSRC units that operate in the 5.9 GHz band. The OBU on the train broadcasts warning messagesprior to approaching the crossing, and the warning lasts untilthe train has cleared the crossing. The warning message is expected to be received at the RSU as well as at the OBU of each of the vehicles near the railway track, thus notifyingall vehicles about the incoming train. The RSU acts as a relay for warning

[1] This work was funded in part by the affiliates of Federal Railroad Administration (FRA) and Wireless@Virginia Tech

messages:it receives the messages from the train and then retransmits themto the vehicles along the road, therefore effectively increasing the coverage range of the warning system. The RSU, thus functions as both a transmitter and a receiver.It is also fitted with two directional antennas, a receive antenna with its axis along the direction of the railway tracks and a transmit antenna with its axis along the roadway. However, since this is a critical safety application, and the failure of the system can result in fatal accidents, it is not wiseto rely entirely on the functioning of RSU. As a result, the RSU provides redundancy and should only be used as a backup, andthe communication system must operate successfully even in the absence, or failure, of the RSU.

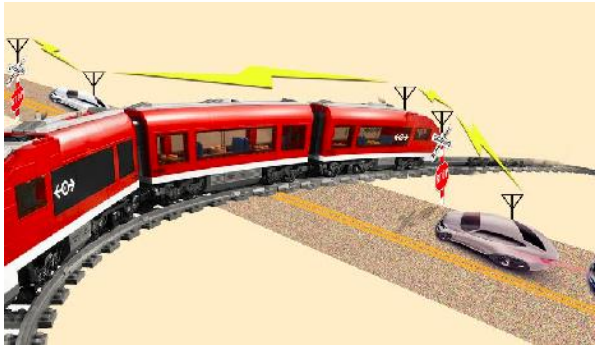

Figure 1   The overall system concept

Thus a direct link between train an vehicle must be established. , In this system, the communication between train and vehicle is no longer on-axis,as the train and the vehicle are approaching each other at some angle (often at approximately 90 degrees) and not along the same roadway as it is for V2V communications. Thus,deploying a directional antenna that focuses its power only along a particular small angle can be unreasonable, since the entire power will be received at only a very small region around the crossing, and vehicles at a slightly larger distance will not receive the warning message at all. That is to say, a vehicle on the roadway will receive the message only when it is very close to the crossing, therefore making the warning system ineffective and increasing the chances of a train-vehicle crash that the system was originally designed to avert.

Despite the advantages of simplicityand low cost, the omnidirectional antenna suffers from one major fault: its gain is very low. Since the transmit power is being spread in all azimuthal directions, including ones in which there are no roads or approaching vehicles, the receive power level will be very small along the roadway and even at the crossing.By the time a sufficiently high power is received at the vehicles, the train could be extremely close to the crossing, once again increasing chances of a train-vehicle collision.

If a directional antenna is used for train-to-vehicle communication, there is a tradeoff among how far along the railroad the directional transmission can reach, the width of the area that antenna beam can cover, and the size of the antenna. The schematicgeographical topology for train-to-vehicle communication is shown in Figure 2. On one hand, the narrower the beam of the transmission, the earlier that the vehicle would receive the warning information. On the other hand, the wider the beam, the larger area that the directional antenna can cover. Those constraintsmust be taken into consideration for design or selection of directional antennas for the warning the system.

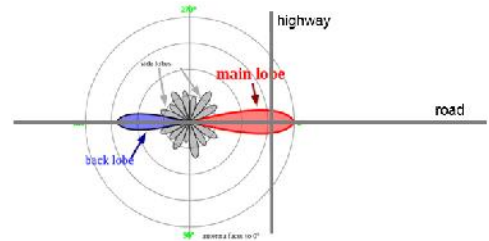

Figure 2   Geographical topology for train-to-vehicle communication

In this paper, we investigate the antenna pattern and the goal is to find the smallest antenna that satisfies the coverage both on the road and the railway.

## II.  RELATED WORKS

Most V2V or V2Icommunicationsystems have been designed based on the DSRC protocol[4, 5].Compared with the omni-directional counterparts, a directional antenna would achieve a better transmit or receive gain in the target direction[7]. Another benefit of using directional antenna is the better immunity from co-channel interference and multi-path fading [8]. TheMobiSteer project [9] studied V2I communication using thedirectional antenna. The same antenna setup is also used in [10] to conduct experiments using commercially available phased-array antennas mounted on cars to show the improvement for a single V2V link in both suburban area and a highway environment.

However, there are very few studies in the literature that address train warning systems using wireless technologies.The only description of such a system found in the literature so far is a train-to-vehicle early warning system which is designed and managed for trains and vehicles near railroad crossings in Australia [6, 11].In those works, DSRC technology has been proved to have the potential (1) to offer a cost effective approach of deploying an ITSthat provides social, economic and environmental benefits, and (2) to help in long term behavior change of drivers and result in an overall improvement in safety consciousness on Australian roads. However, there is no specific study on the directional antenna in that project. In this paper, we focus on the analysis of the directional antenna for train-to-vehicle communication. The goal is to investigate the optimal antenna selection for the railroad crossingsafety application.

## III.  DIRECTIONAL ANTENNA ANALYSIS

In this section, we first present the railroad crossing safety requirement, followed by the discussion on the tradeoff of using directional antenna. After that, an optimal antenna pattern is defined and derived for different geometric scenarios.We then evaluate and discuss the feasibility of an array antenna through measurements on a near field scanner.

## A. Railcross Safety Requirement

According to [12], three factors determine the distance that it takes to stop the vehicle: perception time, reaction distance, and braking distance. The report specifies the average stopping distance on dry, level pavement as shown in Table 1. Considering the highest vehicle speed near the crossing to be 65 mph (105 km/h), the longest stopping distance is thus 105 meters. Because wet pavement may double the braking distance, we consider the worst case that the stopping distances are also doubled, i.e, 210 meters. Although stopping distances in ice or snow are longer, we assume the worst-case distance wouldn't change considering that driving speeds are also reduced in snow.

According to [13], for U.S. railway system, the maximum permissible speed for trains across most of the U.S. is 80 miles/hour (129 km/h), although faster speeds are permissible on a few select tracks. To avoid exceeding the speed limit, most trains limit their maximum speed to 79 mph. Thus the 30-sec notification distance is above 1 kilometer as shown in Table 2.

TABLE 1  AVERAGE STOPPING DISTANCES ON DRY, LEVEL PAVEMENT[12]

| Vehicle Speed (in mph) | Stopping Distance (in meters) |
|---|---|
| 25 | 26 |
| 35 | 42 |
| 45 | 60 |
| 55 | 81 |
| **65** | **105** |

TABLE 2  NOTIFICATION DISTANCES FOR DIFFERENT TRAIN SPEEDS AND REQUIRED NOTIFICATION

| Train Speed (in mph) | Required Notification time (in seconds) | Notification Distance (in meters) |
|---|---|---|
| 79 | 15 | 530 |
| 79 | 20 | 706 |
| 79 | 25 | 883 |
| 79 | 30 | 1059 |

Thus, the scenario we considered is that (1) the receiver sensitivity is -90 dBm, (2) the maximum required transmission range on the railway is 1.06km, (3) the coverage on the road is 210 m on either side of the crossing.

## B. Optimal Antenna Pattern Derivation

In this section, we analyze the antenna pattern in order to find the optimal one with the smallest physical size. The parameters and values used are listed in Table 3.

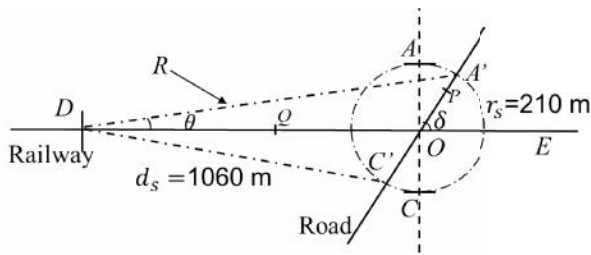

Figure 3  Schematic diagram of the criteria for antenna selection in a any-angle intersection.

The "optimal" antenna in the paper is defined as the smallest antenna which has a radiation pattern that meets the following two criteria.

**Criteria # 1**: When the train is far away (e.g. 30 seconds maximum lead time from Table 2) from the crossing, the gain/pattern should be sufficient to establish connectivity with the DSRC OBU and RSU.

**Criteria # 2**: The gain/pattern should also maintain connectivity as the train passes through the crossing, and should continue to maintain connectivity until the train completely clears the crossing.

TABLE 3  NOTATIONS FOR ANTENNA ANALYSIS

| Symbol | Definition | Value |
|---|---|---|
| $P_t$ | transmitted power | 24 dBm |
| $P_r(d)$ | received power | N/A |
| $G_t$ | transmitter antenna gain | 1 |
| $G_r$ | receiver antenna gain | 1 |
| $d$ | distance between transmitter and crossing | N/A |
| $L$ | system loss | 1 |
| $n$ | path loss exponent | 2 ~ 5 |
| $\lambda$ | wave length in meters | 0.0508 m |
| $v_t$ | the speed of the train | 79 mph |
| $v_v$ | the speed of the vehicle | 65 mph |
| $r$ | the distance from the crossing on the road that the train can cover | N/A |
| $r_s$ | the safe distance from the crossing on the road that the vehicle can have train can cover | 210 meter |
| $R$ | the distance between the antenna of the train and the receiver of the vehicle | N/A |
| $P_{min}$ | minimal receiver sensitivity | - 90 dBm |
| $L_t$ | length of the train | 1.06km [1] |

That is to say, the transmission needs to provide sufficient warning distance along the intersecting road that exceeds driver reaction and braking time no matter the train is far away from or very close to the crossing. Then, an extra constraint need to be added to the problem formulation to make sure the connectivity is maintained until the train completely clears the crossing (assuming a train is configured as a "PULLING", e.g. the antenna is installed at the railway engine, which is at the front of the train). Using the Link Budget Equation, we have

$$P_r(d) = \frac{P_t G_t(\theta) G_r \lambda^2}{(4\pi)^2 R^n L} \geq P_{min} \quad (1)$$

Thus,

$$G_t(\theta) \geq \frac{P_{min}(4\pi)^2 R^n L}{P_t G_r \lambda^2} \quad (2)$$

If we want the antenna to be capable of satisfying both the Criteria 1 and 2, (2) needs to be satisfied. For simplicity, we focus on the case that the length of the train is also 1.1km (same as the safe distance on the railway), but similar methodology can be extended for the general case as well.

We consider the antenna pattern is symmetric around the railway on the azimuthal plane for the any-angle intersection case. The railway and the road are not necessarily perpendicular, and the angle δ ($usually \geq 45°$ considering the practical road geometry) between the railway and the road is not 90°. As shown in Figure 3, to guarantee the safe distance on the road, the received power at point $P$ (on $A'C'$) should be no less than -90 dBm if the train's antenna is at point $D$, which is the edge point of the safe transmission range on the railway. Meanwhile, when the train is approaching the crossing from point $D$ to point $Q$, the received power at all points within $A'C'$ should be no less than -90 dBm in order to either maintain the connectivity with the connected vehicle(s), or notify any unconnected vehicle(s) that enter the safe transmission region on the road while the front of the train is within the safe transmission region on the railway. If the train's antenna is at point $D$, for the point $P$ within $A'O$, we know

$$\tan\theta = \frac{r \cdot \sin(\delta)}{d_s + r \cdot \cos(\delta)} \quad (3)$$

---

[1] Actual train length can be as great as 2 miles (approximately 3.2 km). Similar analysis methodology can be extended for the train with any length.

Thus,

$$r = \frac{d_s \cdot \tan(\theta)}{\sin\delta - \cos(\delta)\cdot\tan(\theta)}, \text{ for } 0 \leq \theta \leq \operatorname{atan}\left(\frac{r_s \cdot \sin\delta}{d_s + r_s \cdot \cos\delta}\right) \quad (4)$$

So, $R = \sqrt{(r \cdot \sin\delta)^2 + (r \cdot \cos\delta + d_s)^2}$, and thus

$$G_t(\theta) \geq \frac{P_{min} 4\pi^2 L}{P_t G_r \lambda^2}((r \cdot \sin\delta)^2 + (r \cdot \cos\delta + d_s)^2)^{\frac{n}{2}} \quad (5)$$

Meanwhile, the received power at point $C'$ is no less than $P_{min}$ if the train's antenna is within DO ($\operatorname{atan}\left(\frac{r_s \cdot \sin\delta}{d_s + r_s \cdot \cos\delta}\right) \leq \theta \leq \pi - \delta$), we have

$$G_t(\theta) \geq \frac{P_{min} 4\pi^2 L}{P_t G_r \lambda^2}((r_s \cdot \sin\delta)^2 \cdot \operatorname{ctan}^2\theta + (r_s \cdot \sin\delta)^2)^{\frac{n}{2}} \quad (6)$$

It is easy to prove that if the inequalities (5) and (6) can be satisfied, the requirement of other cases can also be met, such as the cases that the receive power at point $M$ within $A'O$ should be no less than $P_{min}$ if the train's antenna is at point $D$; the receive power at point $A'$ is no less than $P_{min}$ if the train's antenna is at point $Q$ within $DO$ such that $\operatorname{atan}\left(\frac{r_s \cdot \sin\delta}{d_s + r_s \cdot \cos\delta}\right) \leq \angle A'QE \leq \delta$, etc.

If the gain/pattern should also maintain connectivity as the train passes through the crossing, and should continue to maintain connectivity until the train completely clears the crossing, then the analysis of the backward side beam pattern is the same as the analysis for the forward beam.

Through simulation, the antenna patterns in the azimuthal plane (considering $\delta \geq 45°$) are shown in Figure 4. Note that these are the patterns that meet the minimal requirement of our system for different values of $\delta$.

As shown, other than the slightly larger power requirement (less than 1 dB) around the antenna axis, the perpendicular case (i.e. $\delta = 90°$) covers all the other cases. These results indicate that as long as the perpendicular case is taken into consideration, the antenna design/selection would be enough for railroad grade crossing safety applications.

The "optimal" antenna is considered as the physically smallest one that meets the requirement on the horizontal plane. We assume the antenna gain in the vertical plane is uniformly distributed across $\varphi$ at a particular $\theta$. To derive the antenna size, we integrate the antenna gain across $\theta$ and $\varphi$, we choose the angular resolution to be a very small value Δ, thus

$$\frac{\sum_0^{2\pi}\sum_0^{\varphi_{max}} G_t(\theta,\varphi)}{(2\pi/\Delta)^2} = 1 \quad (7)$$

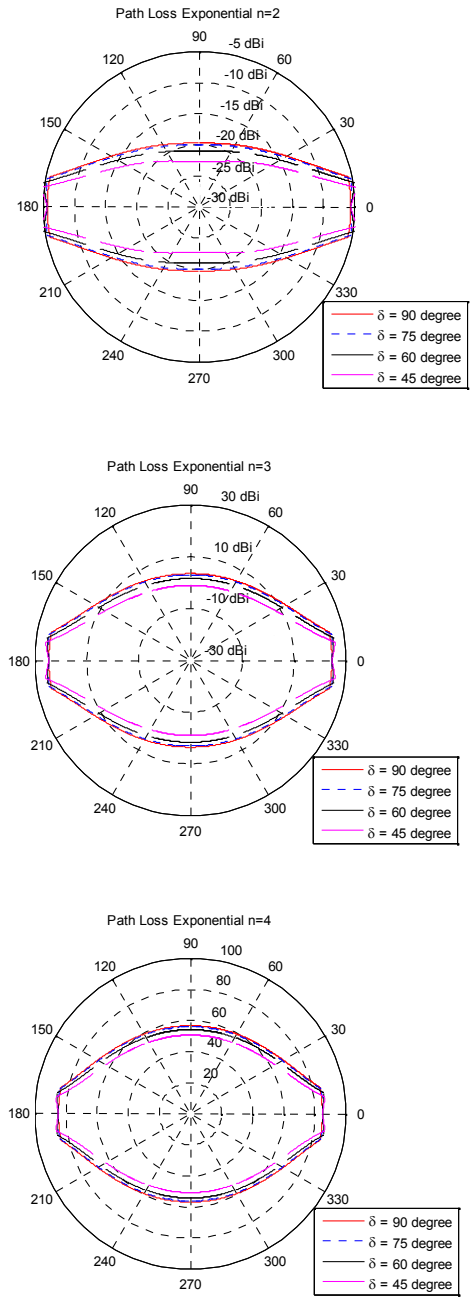

Figure 4  Antenna patterns for different $\delta$.

Then,

$$\varphi_{max} = \frac{(2\pi/\Delta)^2}{\frac{1}{\Delta}\sum_0^{2\pi} G_t\,\theta,\varphi} = \frac{4\pi^2}{\Delta \cdot \sum_0^{2\pi} G_t\,\theta,\varphi} \quad (8)$$

The calculation of the antenna length and width is based on [14]:

$$\text{Length} = \frac{51 \cdot \lambda}{\varphi_{3dB}} \quad (9)$$

$$\text{Width} = \frac{51 \cdot \lambda}{\theta_{3dB}} \quad (10)$$

By varying the required notification time (lead time) from 30 to 15 seconds, the estimated antenna sizes are listed in Table 4 ($n = 3$ and $v_t = 79mp?$). Specifically, we listed the estimated antenna sizes for perpendicular intersection under different path loss factors in Table 5.

TABLE 4  SMALLEST ANTENNA SIZE FOR DIFFERENT REQUIRED NOTIFICATION TIME

| Lead time | Max gain in horizontal plane | Maximum angle in vertical plane | Antenna length | Antenna Width |
|---|---|---|---|---|
| 30 sec | 24.8dBi | 6.4° | 40.7cm | 5.1cm |
| 25 sec | 22.7dBi | 9.0° | 28.8cm | 4.4cm |
| 20 sec | 19.9dBi | 13.9° | 18.7cm | 3.5cm |
| 15 sec | 16.5dBi | 24.3° | 10.6cm | 2.6cm |

TABLE 5  SMALLEST ANTENNA SIZE FOR DIFFERENT PATH LOSS FACTORS

| Lead time | n value | Max gain in horizontal plane | Max angle in vertical plane | Antenna Length | Antenna Width |
|---|---|---|---|---|---|
| 30 sec | 2 | -5.5 dBi | 360° | 0.72 cm | 4.8 cm |
| 15 sec | 2 | -11.0 dBi | 360° | 0.72 cm | 2.5 cm |
| 30 sec | 3 | 24.9 dBi | 6.4° | 40.6 cm | 5.1 cm |
| 15 sec | 3 | 16.5 dBi | 24.3° | 10.7 cm | 2.6 cm |
| 30 sec | 4 | 55.2 dBi | 0.0068° | 379.7 m | 5.3 cm |
| 15 sec | 4 | 44.1 dBi | 0.052° | 50.3 m | 2.7 cm |

*C. Remarks from Numerical Analysis*

The above optimal antenna patterns derived for different path loss exponents provide us with several interesting observations. First, for n=2, the optimal pattern has an on-axis gain of -5.5 dBi. This means that a typical omnidirectional antenna (such as a monopole) is sufficient in meeting all the requirements of our system. Thus, we do not need to use a directional antenna in this case. However, $n = 2$ is rarely observed in practice. For $n = 3$, the optimal antenna requires a main beam gain of about 24.8 dBi. The elevation beamwidth is therefore limited to 6.4 degrees. Most importantly, in order to fulfil all requirements and provide the necessary coverage, we require an antenna with a size of at least 40.7cm by 5.1cm. The antenna size requirement is larger for higher values of path loss exponent.

One possible approach to constructing the optimal antenna is to use arrays of omnidirectional elements to approach an optimal pattern. For example, consider a high-gain omni antenna which has a gain of 15 dBi for all azimuthal angles in the horizontal plane. In order to achieve the required antenna gain of 24.8 dBi (for n=3), a further 10 dB of gain is required. This can be achieved by using an array of 10 such antennas. Then, for an element spacing of λ/2 (e.g. 2.5 cm at 5.9 GHz), the optimal antenna is an array which is effectively 25 cm in length. This length is smaller than the optimal antenna described in this paper (40.7 cm in length), however, the array antenna may not satisfy the required gain at some particular angles, such as the angles around 90 degree.

*D. Antenna Array Measurements*

We conducted measurements to evaluate the performance of the array antenna on a near field scanner in order to determine its gain and directive pattern. The array consists of eight Mobile Mark EC012-5900 vertical antennas with a nominal gain of 12 dBi. The array was fed by an eight way power divider and phase matched cables. The array elements were mounted on a base plate that sets the array spacing. Due to the mechanical limitation of the base plate, the center-to-center distance between adjacent antennas is 3.5 cm.

This antenna array was tested using the planar scanning method. This method was chosen as the linear array conforms best to the planar scan. The planar scan is best for highly directive antennas where the far side lobes are not of primary importance. Since the planar scan is limited in scan distance in both X and Y directions, data is limited to less than 90 degrees from center. In this case the X and Y scans were 58 inches and the distance from the antenna array to the probe (at the scan plane) was 10 wavelengths (about 18 inches). 110 position points were taken in each dimension. This gives a window of about plus or minus 60 degrees with the best data estimated to be at plus or minus 45 degrees. Since the antenna was not scanned around 180 degrees the back lobe is not shown. From symmetry the back lobe should mirror the front lobe. Figure 5 shows the array mounted in the anechoic chamber along with the scanner probe.

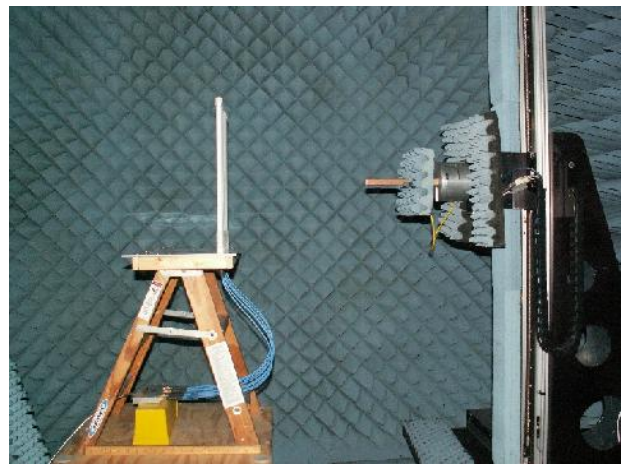

Figure 5  Array antenna mounted in the anechoic chamber along with the scanner probe.

The raw scan results are were processed using a transform to determine the far field patterns. A standard gain horn (Narda 642) was also scanned in order to compute the actual gain by the substitution method. Figure 6 is a polar plot of the array vertical (elevation) pattern. The vertical pattern is centered on the main lobe. There is a slight tilt to the vertical pattern which reduces the gain at the horizon slightly. The vertical pattern corresponds approximately to the vertical pattern of the single element tested previously and shown below in Figure 8. Figure 7 is the horizontal pattern of the array. The horizontal pattern is due primarily to the array factor which is determined by the array geometry rather than the element characteristics. The horizontal pattern is quite good with a beam width of a little more than seven degrees. The array has a gain of 23.5 dBi on the horizontal plane. The overall array gain, however, is slightly higher at the peak which is offset by the vertical pattern. Based on the measurements, it is found that the pattern of the designed array antenna would not perfectly fit the optimal pattern that we derived previously. This is mainly because of the nulls appearing in the array antenna.

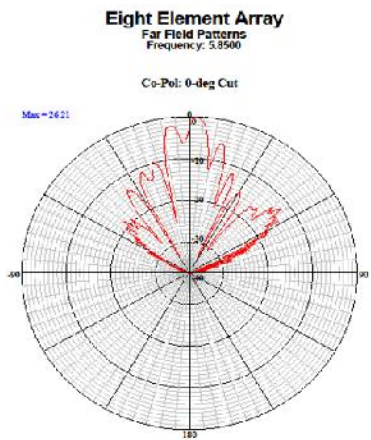

Figure 6    Vertical pattern of the array antenna.

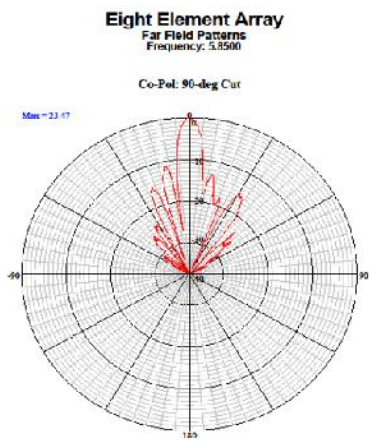

Figure 7    Horizontalpattern of the array antenna.

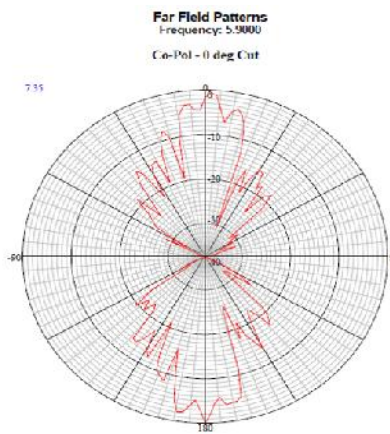

Figure 8    Elevation pattern from single array element.

## IV. CONCLUSION AND FUTURE WORK

In this paper, we analyze the use of directional antennas for railroadcrossing safety applications. First, railroad crossing safety requirements are presented. After describing the tradeoff involved in antenna selection, the optimal antenna is defined and derived for different geometric scenarios.Through the analysis, we find that the consideration of the perpendicular case would be enough to cover the any-angle intersection cases.We have shown that for a maximum transmit power of 24 dBm, the train-to-vehicle safety warning system provides sufficient and reliable warning time with practicable antenna sizes for up to a path loss exponent of 3.We have conducted a feasibility study of one array antenna and demonstrated the advantages and limitations of the designed array antenna.

### REFERENCES

[1] P. Levin, Thomas, Mitchell, Echsner & Proctor, Attorneys at Law. (2012). *Railroad Accident Questions*. Available: http://train-accident-law.com/faq.html

[2] Federal Railroad Administration Office of Safety Analysis. (2013). *FRA Office of Safety Analysis Web Site*. Available: http://safetydata.fra.dot.gov/OfficeofSafety/Default.aspx

[3] J. B. Kenney, "Dedicated Short-Range Communications (DSRC) Standards in the United States," *Proceedings of the IEEE,* vol. 99, pp. 1162-1182, 2011.

[4] C. Hsu, C. Liang, L. Ke, and F. Huang, "Verification of On-Line Vehicle Collision Avoidance Warning System using DSRC," *World Academy of Science, Engineering and Technology,* vol. 55, pp. 377-383, 2009.

[5] Y. Xue, L. Jie, N. F. Vaidya, and Z. Feng, "A vehicle-to-vehicle communication protocol for cooperative collision warning," in *The First Annual International Conference on Mobile and Ubiquitous Systems: Networking and Services*, 2004, pp. 114-123.

[6] J. Singh, A. Desai, F. Acker, S. Ding, S. Prakasamul, A. Rachide*, et al.*, "Cooperative intelligent transport systems to improve safety at level crossing," in *The 12th Global Level Crossing and Trespass Symp., London*, 2012.

[7] R. Ramanathan, "On the performance of ad hoc networks with beamforming antennas," presented at *the Proceedings of the 2nd ACM international symposium on Mobile ad hoc networking& computing, Long Beach, CA, USA*, 2001.

[8] T. Aubrey and P. White, "A comparison of switched pattern diversity antennas," in *43rd IEEE Vehicular Technology Conference*, 1993, pp. 89-92.

[9] V. Navda, A. P. Subramanian, K. Dhanasekaran, A. Timm-Giel, and S. Das, "MobiSteer: using steerable beam directional antenna for

vehicular network access," presented at *the Proceedings of the 5th international conference on Mobile systems, applications and services, San Juan, Puerto Rico*, 2007.

[10] A. P. Subramanian, V. Navda, P. Deshpande, and S. R. Das, "A measurement study of inter-vehicular communication using steerable beam directional antenna," presented at *the Proceedings of the fifth ACM international workshop on Vehicular inter-networking, systems, and applications, San Francisco, California, USA*, 2008.

[11] G. M. Jan Fischer-Wolfarth, *Advanced Microsystems for Automotive Applications*: Springer, 2013.

[12] *Virginia Driver's Manual*. Available: http://www.dmv.state.va.us/webdoc/pdf/dmv39.pdf

[13] "United States Code of Federal regulations."

[14] L. V. Blake, *Antennas*: John Wiley & Sons, 1966.